\documentclass[letterpaper, 10 pt, conference]{ieeeconf}  

\IEEEoverridecommandlockouts                              
\usepackage{graphicx}

\title{\LARGE \bf
A Brief Review on Some Architectures Providing Support for DIFT 
}

\author{Ali Jahanshahi\\
ajaha004@ucr.edu\\
University of California, Riverside
}

\begin{document}

\maketitle
\thispagestyle{empty}
\pagestyle{empty}

\begin{abstract}
Dynamic Information Flow Tracking (DIFT) is a technique to track potential security vulnerabilities in software and hardware systems at run time. The last fifteen years have seen a lot of research work on DIFT, including both hardware-based and software-based implementations for different types of processor architectures. This survey briefly reviews some hardware architectures that provide DIFT support. Starting from introducing different approaches for hardware based DIFT, this survey focuses on integrated/in-core architectures. Protection schemes, including tagging system, tag propagation, and tag checking for each architecture will be discussed. The survey is organized in such a way that it illustrates the evolution of integrated DIFT architectures, each architecture tries to improve the precious proposed architectures generality/versatility weaknesses. However, improving security while providing generality and versatility is kind of trade-offs. This survey compares the architectures from different aspects to show the trade-offs clearer.
\end{abstract}
\section{INTRODUCTION}
Buffer overflows and format strings are the most frequently-exploited program vulnerabilities. These attacks give attackers the ability to change spatially limited yet important memory locations in the vulnerable program’s memory space with malicious code and program pointers. By exploiting these vulnerabilities, a malicious entity will be able to take control of a program and perform any operation with the privileges of the compromised program. Although taking control of a single privileged program grants attackers full access to the system, attacks to hijack any program that has access to sensitive information is considered a serious security threat. Unfortunately, protecting programs by preventing the first step of an attack, which is exploiting program vulnerabilities to modify memory locations, is very difficult.\\
\indent There can be as many, if not more, types of exploits as there are program bugs. Moreover, malicious overwrites cannot be easily identified since vulnerable programs themselves perform the writes. Conventional access controls do not work in this case. As a result, protection schemes which target detection of malicious overwrites have only had limited success. They block only the specific types of exploits they are designed for or they are too restrictive and cannot handle some legitimate programs such as dynamically generated code. To thwart a broad range of security exploits, we can prevent the final step, namely, the unintended use of I/O inputs.\\
\begin{figure*}
  \centering
  \includegraphics[width=0.8\textwidth]{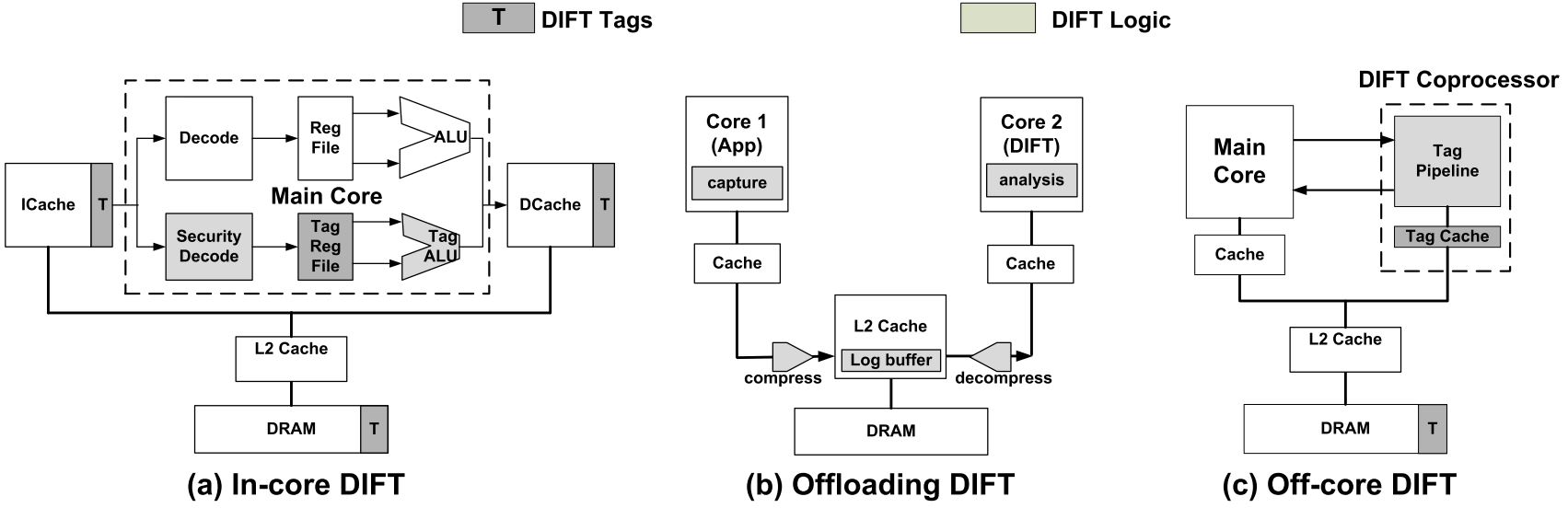}
  \caption{The three design alternatives for DIFT architectures \cite{r2009}}
  \label{fall}
\end{figure*}
\indent Dynamic information flow tracking (DIFT) is a promising technique to detect security attacks on unmodified binaries ranging from buffer overflows to SQL injections. The idea behind DIFT is to tag (taint) untrusted data and track its propagation through the system. DIFT associates a tag with every memory location with a specific granularity. Any new data derived from untrusted data is also tainted using the tag bits. If tainted data is used in a potentially unsafe manner, for instance as a code pointer or as a SQL command, a security exception is immediately raised.\\
\indent The generality of the DIFT model has led to the development of several implementations. To avoid the need for recompilation, most software DIFT systems use dynamic binary translation, which introduces significant overheads ranging from 3x to 37x or even more. Additionally, software DIFT does not work safely with self-modifying and multi-threaded programs. \\
\indent Hardware DIFT systems have been proposed to address these challenges. They make DIFT practical for all user or library executables, including multithreaded and self-modifying code, and even the operating system itself. As Fig.\ref{fall} shows, existing DIFT architectures follow three general approaches\cite{r2009}; Integrated/In-core architectures (Fig.\ref{fall}.a) provide DIFT support within the main pipeline. Most of the proposed DIFT systems follow the integrated approach, which performs tag propagation and checks in the processor pipeline in parallel with regular instruction execution. This approach does not require an additional core for DIFT functionality and introduces no overhead for inter-core coordination. Overall, its performance impact in terms of clock cycles over native execution is minimal. On the other hand, the integrated approach requires significant modifications to the processor core. All pipeline stages must be modified to buffer the tags associated with pending instructions. The register file and first-level caches must be extended to store the tags for data and instructions. Alternatively, a specialized register file or cache that only stores tags and is accessed in parallel with the regular blocks must be introduced in the processor core. Overall, the changes to the processor core are significant and can have a negative impact on design and verification time. Depending on the constraints, the introduction of DIFT may also affect the clock frequency. The high upfront cost and inability to amortize the design complexity over multiple processor designs can deter hardware vendors from adopting this approach.\\
\indent An alternative approach (Fig.\ref{fall}.b) is to offload DIFT functionality to another core in a multi-core chip. The application runs on one core, while a second general-purpose core runs the DIFT analysis on the application trace. The advantage of the offloading approach is that hardware does not need explicit knowledge of DIFT tags or policies. It can also support other types of analyses such as memory profiling and locksets. The core that runs the regular application and the core that runs the DIFT analysis synchronize only on system calls. Nevertheless, the cores must be modified to implement this scheme. The application core is modified to create and compress a trace of the executed instructions. The core must select the events that trigger tracing, pack the proper information (PC, register operands, and memory operands), and compress in hardware. The trace is exchanged using the shared caches (L2 or L3). The security core must decompress the trace using hardware and expose it to software. The most significant drawback of the multi-core approach is that it requires a full general-purpose core for DIFT analysis. Hence, it halves the number of available cores for other programs and doubles the energy consumption due to the application under analysis. The cost of the modifications to each core is also non-trivial, especially for multi-core chips with simple cores. \\
\indent Compared to the multi-core DIFT approach, using a co-processor (Fig.\ref{fall}.C) eliminates the need for a second core for DIFT and does not require changes to the processor and cache hierarchy for trace exchange. The co-processor eliminates the need for any changes to the design, pipeline, or layout of the main core. Hence, there is no impact on design, verification or clock frequency of the main core. Coarse-grained synchronization enables full decoupling between the main core and the co-processor.\\
\indent In this survey, we focus on integrated/in-core architectures. \cite{r2004} proposes and implementation of DIFT on a real processor, \cite{rraksha} tries to improve some problems of \cite{r2004}. \cite{r2018} extends RISC-V processor cores to support hardware dynamic information flow tracking (DIFT). This work is valuable because RISC-V is among the most promising open source hardware solutions. \cite{r2015} tries to extend all the previously mentioned architectures to support software-defined metadata processing which are more general and versatile. Since the architecture proposed in \cite{r2015} is more different from the others, it is investigated separate from other three architectures. And, finally, we compare the architectures from different perspectives.
\section{Protection Scheme}
A DIFT protection scheme relies on three main concepts: \textit{tag initialization}, \textit{tag propagation} and \textit{tag check}. During the tag initialization phase, all data items coming from potentially malicious channels are marked as spurious. In general, a potentially malicious channel is a legitimate I/O communication channel through which malicious inputs may be injected into the application by an attacker. During tag propagation, the processor, depending on the type of instruction that is being executed and on the authenticity of each input operand (as tracked with its corresponding tag), may decide to mark the result of a computation as spurious. This allows the processor to keep track of all information flows generated from spurious inputs at run-time. Finally, during tag checking, if the processor detects that a spurious data item is used in an unsafe manner, it raises a security exception. Thereafter, an exception handler determines whether the use of the spurious value is legitimate or not, i.e. whether the program should resume or terminate. A \textit{security policy} defines the rules that specify how untrusted I/O channels are identified, how dependencies are tracked, and how restrictions on the use of spurious values are applied.
In the following subsections, different aspects of protection scheme that is considered in implementation of three architectural support approaches for DIFT will be described. 
\subsection{Security tags and initialization}
Tags are used to indicate whether the corresponding data block is authentic or spurious. Different approaches has been proposed that extend a one-bit tag to multiple-bit tags in order to distinguish the values or to keep track of multiple security policies concurrently; for example, for values, it may be helpful to distinguish the I/O inputs from the values generated from them. \\
\indent Many architectures support byte granularity memory accesses and I/O. Thus, \cite{r2004} considers one tag bit per byte. In their system, the tags for registers are initialized to be zero at program start-up. Similarly, all memory blocks are initially tagged with zero. In order to initialize the tag of the program data, the authors propose \textit{execution monitor} which is a software module that manages the protection scheme and enforces the security policy. The execution monitor tags the data with one only if they are from a potentially malicious input channel. The other responsibility of execution manager is enforcing the security policy, i.e., in case of a trap generation by processor, the handler checks if the trapped operation is allowed based on the security policy. If the operation is legal, the handler returns to the application. Otherwise, the violation is logged and the application is terminated. Since trap handler is in the OS, security trap handling would be costly. In addition, the execution manager is responsible for tagging potentially malicious inputs. Thus, it should be able to classify the input data accurately so that there is no false positive. The classification process is vague in the proposed architecture and it is a very hard task.\\
\indent The architecture proposed by \cite{rraksha} reasons that since there are a lot of security threats, using just one tag bit is not enough to protect a system. Hence they tried to extend one-bit tag approach. The authors of \cite{rraksha} use 4-bis tags per word. The idea behind using 4-bits tag is to protect against a diverse set of attacks concurrently. In order to initialize tags, security handlers are implemented as shared libraries pre-loaded by the dynamic linker. They claimed that in their architecture, OS ensures that all memory tags are initialized to zero when pages are allocated and that all processes start in trusted mode with register tags cleared. The security handler initializes the policy configuration registers (explained later) and any necessary tags before disabling the trusted mode and transferring control to the application. In order to prevent attackers from compromising the security handler, they introduced another mode orthogonal to user/kernel mode, named trusted/untrusted mode. In trusted mode, tags are only accessible by the security handler and not any other process (even in the kernel mode) can not access/modify them. By using trusted mode there would be no need to change the address space on a security trap which results in fast switching and therefore lower performance overhead in comparison with \cite{r2004}. \\
\indent The proposed implementation in \cite{r2018}, similar to \cite{r2004}, uses 1-bit tag for registers and 1-bit tag for memory bytes, i.e., 4 bits per word. In addition to data registers and memory, they also added a tag bit to PC register. In this implementation, they extended RISC-V ISA with memory and register tagging instructions. Also, library of routines are developed providing APIs to initialize the tags of the data coming from potentially malicious channels at program start-up. \\
\indent  \textbf{Optimizing Tag Memory}: Adding tag bits to all  registers and memory locations leads to a large memory overhead equal to 12.5\%. In order to address this issue, \cite{r2004} introduced \textit{Multi-Granularity Security Tags}. The key idea behind multi-granularity tags is that writing each byte/word separately is not the common case in the applications. In practice, the operating system should maintain two more bits for each page to indicate the type of security tags that the page has. Just after an allocation, a new authentic page holds a per page tag, which is indicated by type value 00. Upon the first store operation with a non-zero security tag to the page, a processor generates an exception for tag allocation. The operating system determines the new granularity of security tags for the page, allocates memory space for the tags, and initializes the tags to be all zero. If the granularity of the store operation is smaller than a quad-word, per-byte security tags are used. Otherwise, per-quad-word tags, which only have 1.6\% overhead, are chosen. If there is a store operation with a small granularity for a page that currently has per-quad-word security tags, the operating system reallocates the space for per-byte tags and initializes them properly. Although this operation may seem expensive, the authors of \cite{}r2004 claimed that it is very rare (happens in less than 1\% of pages).
\indent It is worth mentioning that in all mentioned approaches, the security tags are a part of program state, and should be managed by the operating system accordingly. On a context switch, the tags for registers are saved and restored with the register values. The operating system manages a separate tag space for each process, just as it manages a separate virtual memory space per process.
\subsection{Tag propagation}
A spurious value is the one that may have unexpectedly changed by I/O inputs due to bugs in the program. Once injected, spurious values can again cause unexpected changes to other values through many diﬀerent dependencies. Each of the mentioned approaches categorizes possible dependencies for spurious information flows into different classes. Processors dynamically track spurious information flow by tagging the result of an operation as spurious if it has a dependency on spurious data. To be able to enforce various security policies, the dependencies to be tracked and the way that they are propagating are controlled by a bit vector. \cite{r2004} calls this bit vector Propagation Control Register (PCR), which is set to a proper value by the \textit{execution monitor} based on the specified security policy. \cite{rraksha} names the bit vector Tag Propagation Register (TPR). Since there are four 1-bit tag for enforcing concurrent policies, four TPRs are used in \cite{rraksha}. In both approaches they do not track any form of control dependency. They stated that tracking control dependency is not useful for detecting malicious software attacks considered in their work. Their reason is that for control transfers that can compromise the program, such as register-based jumps, the use of spurious values should simply be checked and stopped.\\
\begin{figure}
  \centering
  \includegraphics[width=0.45\textwidth]{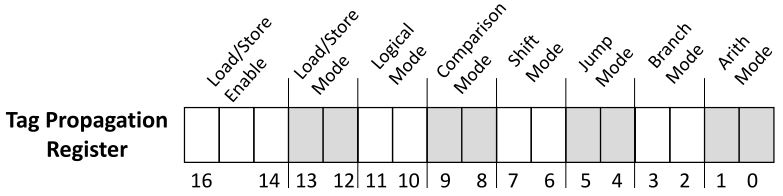}
  \caption{Tag Propagation Register (TPR) proposed in \cite{r2018}}
  \label{ftpr}
\end{figure}
\indent Fig.\ref{ftpr} shows the TPR structure implemented in \cite{r2018}. As it is illustrated, and mentioned before, they considere different dependency classes. In order to propagate the tags, there should be propagation rules for each application which is defined by the programmer through the provided API before starting the program. Table \ref{t2018} shows an example of tag propagation rules specified for securing an application from a specific attack. To make it clearer, in the propagation rule example of Table \ref{t2018}, as an example, the result tag of integer arithmetic operations is logical OR of it's two operands' tag. It means that if any of the operands is spurious, the result of the operation would be tagged as spurious too. Tag propagation rules should be designed carefully because a mistake by the rule designer would be targeted by a security threat.
\begin{table}
\centering
\caption{An Example of Tag Propagation Register Configuration \cite{r2018}}
\label{t2018}
\begin{tabular}{|p{0.13\textwidth}|p{0.03\textwidth}|p{0.25\textwidth}|}
 \hline
\centering
 Field &Value &Rule\\
 \hline
 \hline
 \centering Load/Store Enable &\centering  001 & Source tag enabled\\
 \hline
 \centering Load/Store Mode &\centering 10 & Dest tag = Source tag \\
 \hline
 \centering Logical Mode &\centering 10 & Dest tag = Source1 tag OR Source2 tag\\
 \hline
 \centering Comparison Mode &\centering 00 & No Propagation\\
 \hline
 \centering Shift Mode &\centering 10 & Dest tag = Source1 tag OR Source2 tag\\
 \hline
 \centering Jump Mode &\centering 10 & JAL: New PC = Old PC 
 
 JALR: New PC = Source tag\\
 \hline
 \centering Branch Mode &\centering 00 & No propagation\\
 \hline
 \centering Integer Arith Mode &\centering 10 & Dest tag = Source1 tag OR Source2 tag\\
 \hline
\end{tabular}
\end{table}

\subsection{Checking the Tags and Security Policies}
The tag-check rules restrict the operations that may be performed on tagged data. For example, a tag-check rule may be ``If a tag of a register is set to one then it cannot be used to address the data memory''. Given an instruction, the corresponding check field specifies which operands tags are checked in order to generate a security exception. If the check bit for an operand tag is set to one and the corresponding tag is equal to one, an exception is raised. \cite{r2004} uses Trap Control Register (TCR) that is set by the \textit{execution monitor} based on the security policy. Similarly, as stated before, \cite{rraksha} uses four TCRs to keep track of four security policies concurrently. \\
\indent The security policy defines legitimate uses of I/O values by specifying the untrusted I/O channels, information flows to be tracked (PCR/TPR), tag check conditions (TCR), and software checks on a trap. If the runtime behavior of a program violates the security policy, the program is considered to be compromised. Ideally, the security policy should only allow legitimate operations of the protected program. The policy can be based either on a general invariant that should be followed by almost all programs or on the invariants for a specific application. Also, the restrictions defining the security policy can be based either on where spurious values can be used or on general program behavior.\\
\indent The proposed tag propagation/checking rules in \cite{r2004, rraksha, r2018} are performed by small and simple logical (AND/OR) operations. Thus, tag propagation and checking logic are embedded in execution stage (as a part of ALU) that result in low power and area overhead. Since tag propagation and check is performed in parallel with the actual operation, their logic would not affect the critical path.

\section{Software-Defined Metadata Processing}
\begin{figure}
\centering
\includegraphics[width=0.45\textwidth]{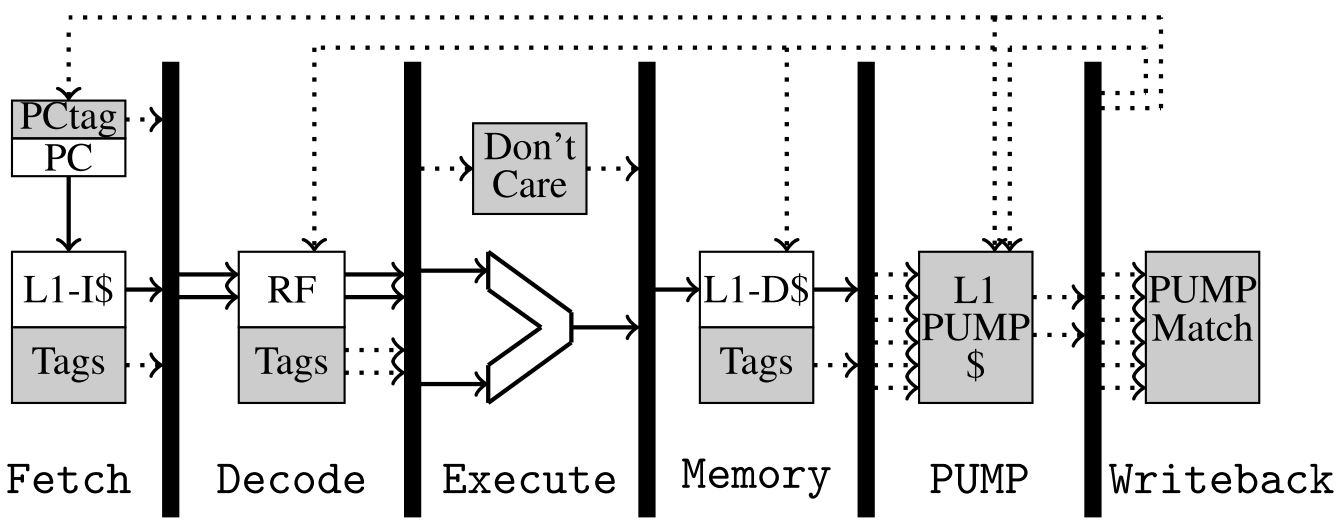}
\caption{the PUMP processor Pipeline}
\label{fpump}
\end{figure}

\indent The designs that were mentioned in previous section have made the hardware metadata computation configurable but one of their main limitation is that they use few bits to represent metadata and only support a limited class of policies. However, fully securing systems will require more than memory safety and isolation. Beside limited bits for the tag, tag propagation rules for every operation (O) in previous approaches only has to do with the operands' tag (OP1/OP2/MR) that produce the result tag (R).
\begin{table}
\centering
\caption{Tag Propagation Rule Format in PUMP and other Architectures}
\label{ttpr}
\begin{tabular}{|p{0.09\textwidth}|p{0.31\textwidth}|}
\hline
\centering Arch& Propagation Rule \\
 \hline
 \hline
\centering \cite{r2004, rraksha, r2018} &O : (OP1, OP2, MR) $\Rightarrow$ (R) \\
 \hline
 \centering PUMP &O : (PC, CI, OP1, OP2, MR) $\Rightarrow$ ($PC_{New}$, R) \\
\hline
\end{tabular}
\end{table}
On the other side of the arena, attacks are rapidly evolving to exploit any forms of vulnerability. Thus, a flexible security architecture that can be quickly adapted to this ever-changing security world is needed. This can be achieved by supporting more general policies and more flexible tag propagation rules in hardware. As shown in Table \ref{ttpr}, a general propagation rule for any operation (O) should be able to decide that what the tag on the program counter in the next machine state ($PC_{New}$) and the tag on the instruction’s result (R) should be if the current tag on the program counter is PC, the tag on the current instruction is CI, the tags on its input operands (if any) are OP1 and OP2, and the tag on the memory location (in case of load/store) is MR. In order to provide support for extensible, software-defined metadata processing with low overhead, \cite{r2015} introduces an architectural model named the Programmable Unit for Metadata Processing (PUMP). PUMP is an extension to a conventional RISC processor (Alpha) and an in-order implementation with a 5-stage pipeline.  \\
\indent Fig. \ref{fpump} shows the processor pipeline of the PUMP. As it is shown, tags are added to all memory elements. The idea behind PUMP is that, for checking the security policy of each operation, a software policy function should be invoke. Invoking a policy function to check whether an operation is allowed or not costs a lot of performance and power overhead because for each operation there should be a context switch to policy function so that it can check if that operation is allowed or not. Statistics shows that security policies are limited and have high locality. Thus, they solved this problem by using a \textit{Rule Cache} added to a stage before committing the executed operation. The operations are checked if they are allowed to commit only by looking up \textit{Rule Cache}. It will be explained the scenario of a rule miss in \textit{Rule Cache} in details.\\
\begin{table*}[!ht]
\centering
\caption{Summary of the Reviewed DIFT Architectures}
\label{tall}
\begin{tabular}{|p{0.035\textwidth}|p{0.07\textwidth}|p{0.1\textwidth}|p{0.18\textwidth}|p{0.07\textwidth}|p{0.08\textwidth}|p{0.08\textwidth}|p{0.18\textwidth}|}
\hline
 Arch & 
 Tag size & 
 Trap Handler &
 Overhead &
 New Instruction &
 Adding Trust Mode &
 \# of Policies &
 Versatility\\
\hline
\hline
\cite{r2004} & 1/byte & OS Level & 1.4\% memory, 1.1\% perf & Yes & No & One & Low, I/O tagged by OS\\
\hline
\cite{rraksha} & 4/word & User Level & 7.17\% logic, 1.34\% perf & Yes & Yes & Four & I/O tagged by sysAdmin\\
\hline
\cite{r2018} & 1/byte & -- & 12.5\% memory, 0.83\% logic & Yes & No & One & High, I/O tagged by the user \\
\hline
\cite{r2015} & word/word & User Level & 110\% area, \textless10\% perf & Yes & Yes & Unbounded & High, I/O tagged by the user\\
\hline
\end{tabular}
\end{table*}
\textbf{Metadata Tags}: Pointer-sized tags are used in this architecture to make the policy writer able to point to any memory location for addressing tags. This notion of software-defined metadata and its representation as a pointer-sized tag extends previous tagging approaches such as \cite{r2004,rraksha, r2018}, where only a few bits are used for tags with fixed interpretation. In order to make the previous approaches extensible, in this architecture, every word is associated with a pointer-sized tag. Thus, the tag is large enough to indirect to a data structure in memory which leads to supporting unbounded metadata. All tag manipulation is defined and implemented with PUMP rules and stored in \textit{Rule Cache}. In this architecture, even worse than the previous approaches, tag memory overhead is a lot (around 2x). In practice, they did not allotted pointer-sized tags. Instead, the authors used 12-bits tag for L1 cache and 16-bits tag for L2 cache. In order to make tag bits consistent when moving the data between caches, they used a translation approach. They mentioned that other methods like \textit{Multi-Granularity Security Tags} can also be used. \\
\textbf{Propagation Rules}: The rules format used in this architecture allows two output tags to be computed from up to five input tags, which makes it a more powerful and versatile architecture. For some security policies, there is no need for all five input tag, and should not be considered. In Fig. \ref{fpump}, \textit{Don't Care} component is responsible for removing any of PC, CI, OP1, OP2, MR in \textit{Rule Cache} lookup if they are not needed for a specific operation (O). By doing so, L1 PUMP cache would have less entropy that results in higher hit rate. \\
\textbf{Rule cache} is added to architecture to support single-cycle common-case computation on metadata. It means, as long as we hit in this cache, any extra cycles would not be added to the execution of the program. Rules are not generated in this cache, this cache is just to look them up. In case of a rule miss, a user-level \textit{miss handler} routine will be invoked to generate a new rule base on the operation, its operands, PC, and CI. Once the rule generated and placed in the \textit{Rule cache}, it will be used in the next accesses. Obviously, the hit/miss rate of this cache depends on how good the user/policy write write the policies, the higher the locality of a policy, the higher hit rate, and the lower performance overhead. \\
\begin{table}
\centering
\begin{tabular}{|p{0.01\textwidth}|p{0.42\textwidth}|}
\hline
\centering  & \textbf{Algorithm 1} N-Policy Miss Handler \\
 \hline
 \hline
1: & for i=1 to N do \\
2: & -- M[i] = \{op, PC[i], CI [i], OP1[i], OP2[i], MR[i]\} \\
3: & -- \{pc[i] , res[i]\} = policy[i](M[i]) \\
4: & $PC_{new}$ = canonicalize([pc[1], pc[2], ..., pc[N]]) \\
5: & R = canonicalize([res[1], res[2], ..., res[N]])\\
 \hline
\end{tabular}
\end{table}
\textbf{Miss Handler}: will be invoked in case of a rule miss in the rules' last level cache. In order to serve the miss, the processor state will be stored in specific registers (added to the processor for this purpose), and the control of the processor goes to the policy function to decide if the operation is allowed or not. In case the operation which caused the miss is a valid operation, new rule will be generated and installed in the \textit{Rule Cache}. In miss handler mode, \textit{Rule cache} is ignored. Similar to \cite{rraksha}, a new mode is introduced in order to provide isolation between the privileged miss handler and the rest of the system software and user code. This mode which is called \textit{miss handler operational mode} is controlled by a bit in the processor state. Miss handler is designed in such a way that multiple policies can be enforced simultaneously. Alg.1 shows the general behavior of the composite miss handler for any N policies. \textit{canonicalize} function is utilized to minimize the number of distinct tags (and hence rules) by using a single tag for logically equivalent metadata. This helps increasing \textit{Rule Cache} hit rate, which has direct effect on the performance.
\section{COMPARISON}
\indent Adding DIFT support to both hardware and software introduces overheads. Table \ref{tall} summarizes the different aspects of the proposed architectures that support DIFT. In Table \ref{tall}, trap handler field means that the security traps are addressed in which level of the system. Also Versatility means that how convenient it is for an application developer to decide which data to tag as potentially malicious data which should be tracked.
\section{CONCLUSION}
\indent In this survey, we briefly reviewed four architectures that provided DIFT support. Adding DIFT support to processors introduces some area/power overhead. As expected, the more secure, general, and versatile the architecture be, the more overhead will be added to the processor. 
\addtolength{\textheight}{-12cm}   



\begin{thebibliography}{99}
\bibitem{r2004} G. E. Suh, J. W. Lee, D. Zhang, and S. Devadas. Secure program execution via dynamic information flow tracking. In International Conference on Architectural Support for Programming Languages and Operating Systems, pages 85–96, 2004.
\bibitem{rraksha} M. Dalton, H. Kannan, and C. Kozyrakis. Raksha: a flexible information flow architecture for software security. In International Symposium on Computer Architecture (ISCA), pages 482–493, 2007.
\bibitem{r2018} C. Palmiero, G. Di Guglielmo, L. Lavagno, and L. P. Carloni, “Design and implementation of a dynamic information flow tracking architecture to secure a RISC-V core for IoT applications,” in Proc. IEEE High Perform. Extreme Comput. Conf. (HPEC), 2018, pp. 1–7
\bibitem{r2015} Udit Dhawan, Catalin Hritcu, Raphael Rubin, Nikos Vasilakis, Silviu Chiricescu, Jonathan M Smith, Thomas F Knight Jr, Benjamin C Pierce, and Andre DeHon. 2015. Architectural support for software-defined metadata processing. ACM SIGARCH Computer Architecture News (2015).
\bibitem{r2009} H. Kannan, M. Dalton, and C. Kozyrakis. Decoupling dynamic information flow tracking with a dedicated coprocessor. In Proc. of the IEEE/IFIP International Conference on Dependable Systems and Networks, pages 105–114, 2009.
\end{thebibliography}
\end{document}